\documentclass{aa} \usepackage{txfonts,graphicx}
\usepackage{natbib}
\begin{document}
\title{Kinematics of diffuse ionized gas in the disk halo interface of
NGC 891 from Fabry-P\'erot observations}

\author{P. Kamphuis \inst{1} \and R.F. Peletier \inst{1} \and
R.-J. Dettmar \inst{2} \and J.M. van der Hulst \inst{1} \and P.C. van
der Kruit \inst{1}\and R. J. Allen \inst{3} } \institute{Kapteyn
Astronomical Institute,University of Groningen, Postbus 800, 9700 AV
Groningen, the Netherlands \and Astronomisches Institut,
Ruhr-Universität Bochum, Universit\"atsstrasse 150, D-44780 Bochum,
Germany \and Space Telescope Science Institute, 3700 San Martin Drive,
Baltimore, MD 21218, USA} \abstract{The properties of the gas in halos
of galaxies constrain global models of the interstellar
medium. Kinematical information is of particular interest since it is
a clue to the origin of the gas.}{ Here we report observations of the
kinematics of the thick layer of the diffuse ionized gas in NGC\,891
in order to determine the rotation curve of the halo gas.}  {We have
obtained a Fabry-P\'erot data cube in H$\alpha$ to measure the
kinematics of the halo gas with angular resolution much higher than
obtained from HI 21\,cm observations. The data cube was obtained with
the TAURUS II spectrograph at the WHT on La Palma. The velocity
information of the diffuse ionized gas extracted from the data cube is
compared to model distributions to constrain the distribution of the
gas and in particular the halo rotation curve.}  {The best fit model
has a central attenuation $\tau_{{\rm H \alpha}}=6$, a dust scale
length of 8.1 kpc, an ionized gas scale length of 5.0 kpc. Above the
plane the rotation curve lags with a vertical gradient of -18.8 km
s$^{-1}$ kpc$^{-1}$. We find that the scale length of the H$\alpha$
must be between 2.5 and 6.5 kpc.  Furthermore we find evidence that
the rotation curve above the plane rises less steeply than in the
plane. This is all in agreement  with the velocities measured in the
HI. }{}  \maketitle

\section{Introduction}\label{intro}
Over the last decade, diffuse ionized gas (DIG) in the halos of spiral
galaxies has been identified as an important constituent of the
interstellar medium (ISM). The detection of an extended layer of DIG
in NGC 891 ($z_{{\rm NW}}$=0.5 kpc, $z_{{\rm SE}}$=0.3 kpc,
\cite{1990A&A...232L..15D}) , which was found to be similar to the
extended layer of DIG, or Reynolds layer, \citep{1990IAUS..139..157R}
of the Milky way \citep{1990A&A...232L..15D,1990ApJ...352L...1R}, was
followed by several H$\alpha$ imaging searches. By now, many  results
on `normal' (i.e., excluding nuclear starbursts) edge-on galaxies have
been published
\citep{1992FCPh...15..143D,1992ApJ...396...97R,1994ApJ...427..160P,1996ApJ...462..712R,2003A&A...406..505R}. \cite{2003A&A...406..505R}
cataloged 74 galaxies and found about 40\% to have extraplanar diffuse
ionized gas (eDIG). In those objects showing H$\alpha$ emission from
the halo, a wide range of the spatial distributions have been found,
from thick layers with filaments and bubbles (NGC 4631, NGC 5775)
\citep{1990A&A...232L..15D,1990ApJ...352L...1R,1994ApJ...423..190P,1999ApJ...522..669H,2003ApJS..148..383M}
to individual filaments and isolated plumes (e.g., UGC
12281)\citep{2003A&A...406..505R}. For only a few of them there is
evidence for widespread DIG in the halo comparable to that in NGC
891. In this galaxy the DIG is distributed in long filaments and
bubbles of ionized gas embedded in a smooth background.\\
\hspace*{0.5 cm}Since its emission line spectrum is rather easily
accessible by optical imaging and spectroscopy,  the DIG component is
an important tracer of the ISM halo in other galaxies. This is true
particularly since most other tracers, such as radio continuum from
cosmic rays or X-rays from hot plasma, cannot be observed  either with
comparable angular resolution or with sufficient sensitivity.\\
\hspace*{0.5 cm}The origin and ionization source of the DIG component
is still under debate and gives important constraints for models of
the ISM in general and on the large-scale exchange of matter between
disk and halo in particular  \citep[e.g.,
][]{1992FCPh...15..143D,1997ApJ...474..129R}.\\
\hspace*{0.5 cm}Theorists describe the disk-halo interaction by means
of galactic fountains
\citep{1976ApJ...205..762S,1980ApJ...236..577B,2005A&A...436..585D},
chimneys \citep{1989ApJ...345..372N}, and galactic winds
\citep{1991A&A...245...79B,1999A&A...347..650B}. Possible models
trying to explain gaseous galaxy halos as a consequence of stellar
feedback therefore depend on many factors, such as supernova rates,
galaxy mass, magnetic fields and the vertical structure of the ISM.\\
\hspace*{0.5 cm}NGC 891 and NGC 4631 are two galaxies with extensively
studied ISM halos. Both of them not only show prominent thick layers
of DIG, they also have extended radio continuum, HI, and X-ray
halos. The spatial correlation of radio continuum emission, indicative
of cosmic rays in a magnetic field found in a thick disk, and
extra-planar DIG has been discussed for NGC 891 in detail
\citep{1992FCPh...15..143D,1994A&A...290..384D}.\\
\hspace*{0.5 cm}If the DIG and other components of the ISM in the halo
are due to dynamical processes, important information on its origin
and ionization could come from kinematic studies. In the case of NGC
891 a first study was made by \cite{1991ApJ...374..507K}. Subsequent
studies show that there is evidence for peculiar velocities of DIG.
\cite{1994ApJ...423..190P} find a maximum difference with the HI
rotation curve $\delta v_{{\rm max}}$=40 km s$^{-1}$;
\cite{1997ApJ...474..129R} retrieves a difference in the observed mean
velocity of 30 km s$^{-1}$ between velocities at z=1 kpc and z=4.5
kpc. Also in the  HI peculiar velocities are observed
\citep{2005ASPC..331..239F,1997ApJ...491..140S}. For both components,
a deviation from corotation is observed on scales of 2 kpc above the
disk in the sense that the gas rotates more slowly than expected.
This ``lagging'' has been found to have a gradient of $dV_{{\rm
rot}}/dz$=-15 km s$^{-1}$kpc$^{-1}$ in HI
\citep{2005ASPC..331..239F}. Recent SPARSE-PAK observations
\citep{2006ApJ...647.1018H} show a similar result for H$\alpha$.\\
\hspace*{0.5 cm} In order to understand this lagging, hydrostatic
models have been investigated. These models are able to reproduce the
lag of the halo of NGC 891 in HI \citep{2006A&A...446...61B}. However,
the stability of these models remains unresolved. A different approach
to understanding the lag of halos are ballistic models
\citep{2002ApJ...578...98C,2006MNRAS.366..449F}.
\cite{2006MNRAS.366..449F} are able to reproduce the vertical HI
distributions of NGC 891 and NGC 2403 this way. However, their model
fails in two important aspects: (1) they do not reproduce the right
gradient in rotation velocity; (2)  for NGC 2403 they predict a
general outflow where an inflow is observed.\\
\hspace*{0.5 cm} It is clear that improved data on the detailed
kinematics of the extra-planar DIG would be very useful to a further
physical understanding of the phenomenon.\\
\hspace*{0.5 cm}Here we present a full velocity cube for the DIG in
NGC 891 from observations with the TAURUS II imaging Fabry-P\'erot
spectrograph. As a byproduct, we obtain a very clean map of the
H$\alpha$ distribution. NGC 891 has a systemic velocity of 528 km
s$^{-1}$ (RC3) and we assume a distance of 9.5 Mpc
\citep{1981A&A....95..116V}. At this distance 1 arcmin corresponds to
2.8 kpc physical size. We present the observations in $\S$ \ref{obs}
and the data reduction steps in $\S$ \ref{data}. $\S$ \ref{result}
will show the results that can be obtained by rebinning the data. In
$\S$ \ref{models} we will present models for the gas distribution and
these models will be discussed and compared to the data in $\S$
\ref{disk}. We will summarize and conclude in $\S$ \ref{summary}.

\section{Observations}\label{obs}
\begin{figure}[Htp]
\centering \includegraphics[width=8cm]{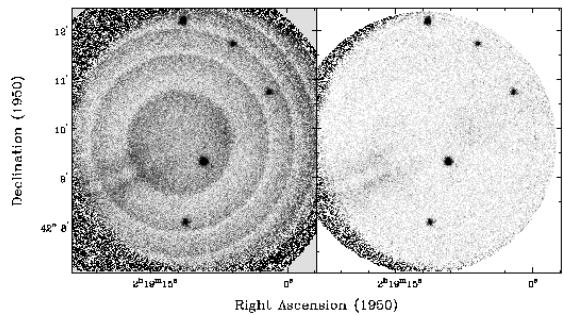}
\caption{This channel of original wavelength calibrated data cube at
v=674 km s$^{-1}$ demonstrates the contamination by night sky lines
with varying intensity (left). The effect of removing a model for the
night sky contribution from averaging in rings is seen on the
right.}\label{fig1}
\end{figure}
\begin{figure}[tp]
\centering \includegraphics[angle=0,width=8cm]{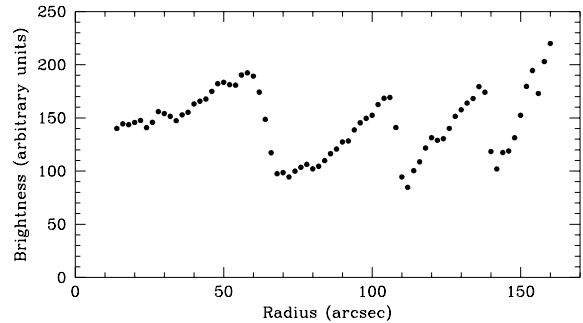}
\caption{Model for the night sky contribution obtained by averaging in
rings as used for the channel shown in Fig. 1.}\label{fig2}
\end{figure}
The data were obtained during two nights in November 1992 with the
TAURUS II imaging Fabry-P\'erot spectrograph at the William-Herschel
Telescope on La Palma. The attached EEV3 CCD detector with a pixel
size of 22.5 $\mu$m provided an image scale of 1.''04/pixel with a
binning by 2. An interference filter with central wavelength at
$\lambda=$657.7 nm and a bandpass of 1.5 nm was used for order
separation. The field of view was restricted to 5.6'$\times$5.6' due
to blocking by the prefilter. Two slightly overlapping fields in NGC
891 were therefore observed to cover the inner $\sim$ 10'. The
North-East field was located at RA$=2^h 22^m 36.8^s$, DEC$=42^{\circ}
22' 32.2"$ (J2000) and the South-West field at RA$=2^h22^m26.7^s$,
DEC$=42^{\circ}18' 5.8"$ (J2000). The observations were taken under
non-photometric conditions, with poor seeing of typically 2
arcsec. For object exposures 75 etalon steps, with a step size of
$\Delta \lambda=$0.02742 nm,  were used with an integration time of 70
s each. Full data cubes were taken at the beginning of each night with
a flat field lamp, and during the night several cubes were taken using
the CuNe lamp to allow us to determine the wavelength dependence in
each channel.  \\
\begin{table}[tbp]
\begin{tabular}{lllll}
\hline Date&Run-No:&Field&UT/Start&Airmass/Start\\ \hline Nov 16/17
1992&3559&NE&19:29   &1.03\\  &3564&SW&00:44 &1.05\\ Nov 17/18
1992&3581&NE&21:46  &1.04\\  &3583&NE&01:09 &1.10\\ \hline
\end{tabular}
\caption{Log of the observations} \label{logtab}
\end{table}
\section{Data Reduction}\label{data}
A slight gradient in the bias level of 1 ADU across the field was
removed in all data cubes because at the amplifier setting this
accounted for a 10$\sigma$ signal with respect to the read-out
noise. The IRAF package was used for these first data-processing steps
including flatfielding and cosmic-ray filtering. For all further
processing steps we made use of the GIPSY package, which is better
suited for these kinds of data cubes.\\
\hspace*{0.5 cm}Cuts in the velocity direction averaged over areas of
the night sky showed significant variations from channel to channel,
up to 20\% in the night sky contribution with a systemic jigsaw
pattern. This pattern is due to the stepping pattern used for scanning
over the $\sim$ 2 hrs of observation for each on-source
cube. Significant brightness changes of two bright night sky lines
were not correlated to these changes in the sky background. We
corrected for the variations of the night sky brightness by
subtracting a constant determined for each channel.\\
\hspace*{0.5 cm}The phase calibration was obtained by fitting a model
to the scans of the CuNe lamp; this way the step size was determined
to be 0.02748 nm. Rebinning the object data cubes with the appropriate
model resulted in four complete data cubes on the object, three for
the field in the North-East and one for that in the  South-West.\\
\hspace*{0.5 cm}The rebinned images still contained the night sky
lines. The strong OH night sky lines at $\lambda=6568.78$ $\AA$ and
$\lambda=6577.28$ $\AA$ were used to establish the absolute wavelength
calibration. They also provide a check on the channel step size. The
$\Delta \lambda=$0.02742 nm determined this way is in excellent
agreement with the determination from the afore-mentioned calibration
cube and corresponds to 12.5 km s$^{-1}$ at H$\alpha$ (for
Fabry-P\'erot data reduction techniques see
\cite{1989AJ.....98..723B,2002MNRAS.329..759J}).  The profile of the
night sky lines also provides information on the spectral resolution,
which was determined to be 40.7  km s$^{-1}$(FWHM).  The formal errors
of the Gaussian fits to the OH-night sky lines allowed us to estimate
the error of the wavelength scale to be less than 6 km s$^{-1}$.  A
correction to the observed velocity of -4.2 km s$^{-1}$ was needed to
obtain the heliocentric velocity.\\
\hspace*{0.5 cm}At this stage, one remaining problem was caused by the
redistribution of the varying line intensity of the night sky lines
during the integration of the observed cube into a wavelength
cube. Rebinning of the lines into the appropriate wavelength channels
resulted in a strong pattern of rings. In the left-hand panel of
Fig. \ref{fig1}, we show this effect for the worst case.  To overcome
this artifact for all affected channels the rings of the line emission
were cut out interactively in areas well separated from the galaxy by
using the GIPSY routine BLOT. The result was integrated using the
routine ELLINT using the center as determined from the phase
calibration and the mean value in individual rings was used as a
model.  Such a ring model is given in Fig. \ref{fig2}. The subtraction
of the model resulted  in general in a satisfactory reduction of the
artifact as demonstrated in the right hand panel of Fig. 1. Typically,
the resulting residuals are smaller than the noise level of the night
sky. This can be judged from the right panel in Fig. 1. However, some
of the channels showed residual larger than the noise level of the
night sky. These residual rings were masked manually.\\
\hspace*{0.5 cm}These cleaned and wavelength-calibrated data cubes
covered a velocity range of 940 km s$^{-1}$, sufficiently large to
provide us with a scaled sum of continuum channels to correct for the
continuum. This continuum correction also removed all ghost images
from internal reflections of the instrument.\\
\hspace*{0.5 cm} We flux calibrated the observations by comparing 7
HII regions in the integrated velocity map with the calibrated
H$\alpha$ image from   \cite{2006ApJ...647.1018H}.  4 of these regions
were located in the NE pointing and 3 in SW pointing. We estimate the
uncertainty of this calibration to be $\sim$ 10\%.\\
\hspace*{0.5 cm}For the merging of all data and for comparison with
other data sets, in particular the HI map provided by
\cite{2005ASPC..331..239F}, astrometry was performed.  Since the two
fields do not sufficiently overlap, we used five stars with positions
obtained from DSS to re-grid the two fields into a common map. The
astrometric accuracy from the fits to the stars is $\sim$ 2 arcsec.
Finally the data cubes were combined into one cube, rotated by 42
degrees in position angle to be oriented along the major axis and cut
back to 42 channels to cover the velocity spread in NGC 891. The noise
in a channel in the fully reduced and calibrated cube is 1.2$\times
10^{-18}$  erg s$^{-1}$ cm$^{-2}$ arcsec$^{-2}$ in the NE pointing and
1.4$\times 10^{-18}$  erg s$^{-1}$ cm$^{-2}$ arcsec$^{-2}$ in the SW.

\section{Results}\label{result}
 
For the following analysis, in order to obtain a better S/N, the data
were binned. In order maintain resolution in higher emission parts,
this was done in such a way that the length and width of a bin
increases exponentially as the distance to the major and minor axis
increases. For the North-East side of the galaxy no binning was
applied when the S/N in a pixel was $\geq$ 4. The channel at the
systemic velocity ($v_{{\rm sys}}=528$ km s$^{-1}$) was set to 0 km
s$^{-1}$ and all velocities given are offsets from this channel. The
central position was determined by eye in several Palomar Sky Survey
and 2MASS images to be RA$=2^h 22^m 33.0^s$, DEC$=42^{\circ} 20'
51.5"$ and set to 0 in the images. From the scatter of the central
position in the different bands we determine the error to be less than
2 arcsec. Notice that this value differs from the best determined
position given by the NASA Extra-galactic Database by almost 6 arcsec
in declination.\\
\hspace*{0.5 cm}For display purposes the figures shown in this paper
come from a cube which was masked so that only regions with signal are
shown. The mask was constructed by smoothing the original binned cube
with a Gaussian of 4 arcsec FWHM, which was  cut at the 1$\sigma$
level.
\subsection {Channel maps}\label{channel}

\begin{figure*}[tbp]
\centering \includegraphics[width=18cm]{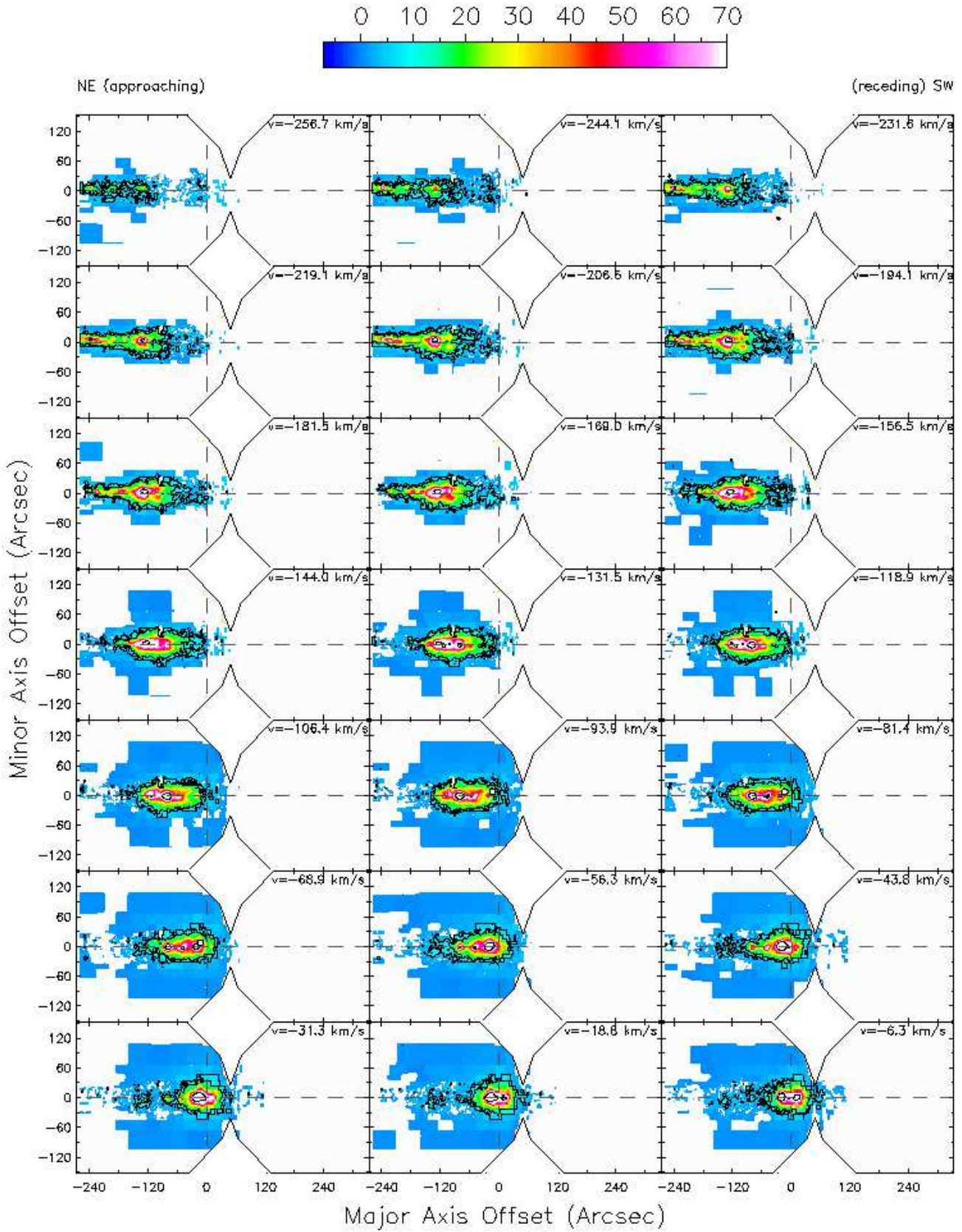}
\caption{The velocity channel maps of the approaching side for the
binned H$\alpha$ of NGC 891. The contours are at 25.3, 51, 510 $\times
10^{-19}$ erg s$^{-1}$ cm$^{-2}$ arcsec$^{-2}$. The horizontal and
vertical dashed lines indicate the major and minor axis
respectively. }\label{fig3a}
\end{figure*}

\begin{figure*}[tbp]
\centering \includegraphics[width=18cm]{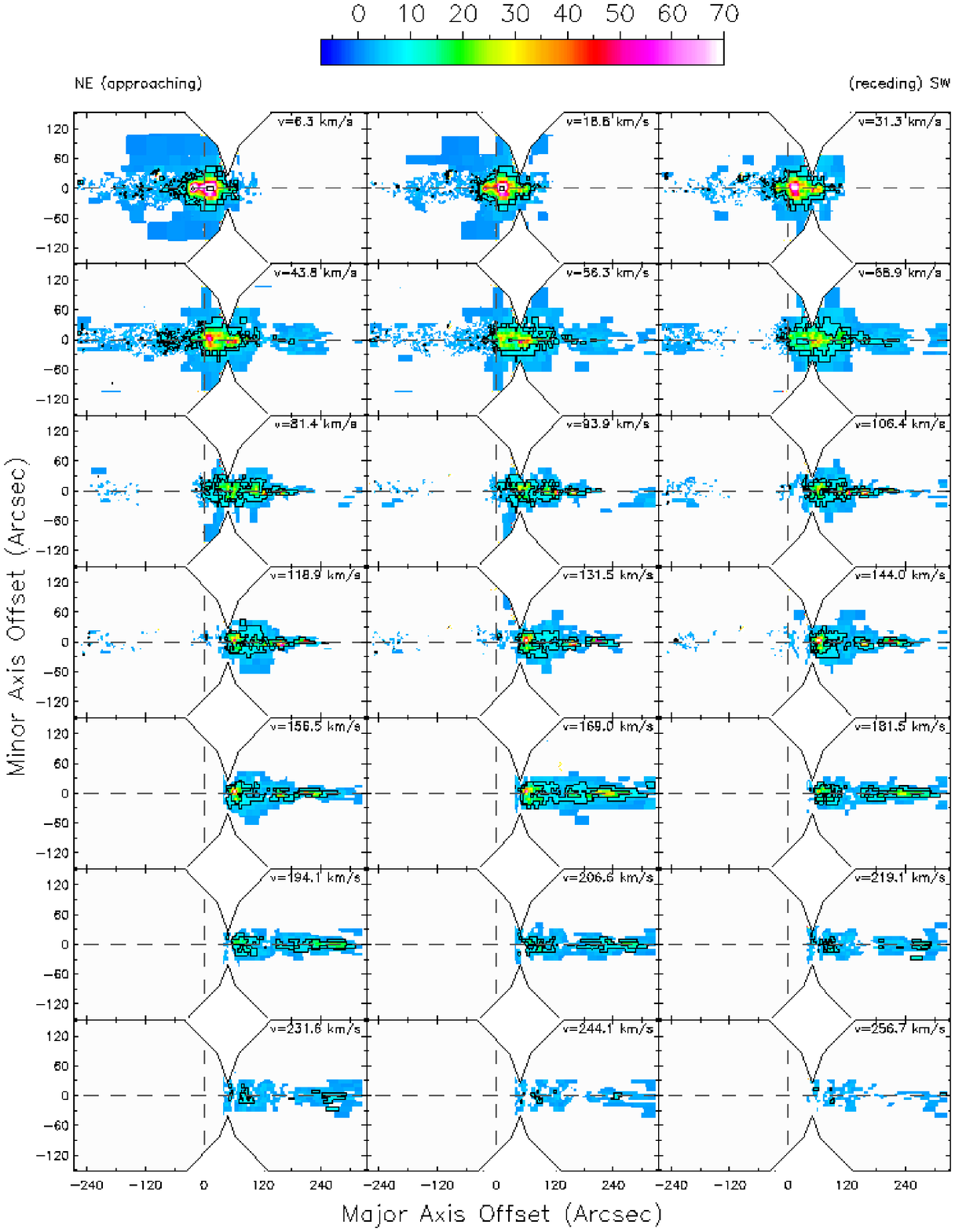}
\caption{Same as fig. \ref{fig3a}, for the receding
side. }\label{fig3b}
\end{figure*}

In Figs. \ref{fig3a} and \ref{fig3b} we give the resulting channel
maps of H$\alpha$ emission with a velocity step size of $\Delta$v=12.5
km s$^{-1}$. In the following figures, the NE part of the galaxy is to
the left; this is also the approaching side of the galaxy. Data are
missing in small wedges along the minor axis, as we had underestimated
the vignetting of the field  when the required overlap of the fields
was determined. It is noteworthy that a thick component in the
H$\alpha$ emission is already visible in individual channel maps. This
sudden thickening of the H$\alpha$-emitting gas layer was reported
before from H$\alpha$ imaging
\citep{1990ApJ...352L...1R,1990A&A...232L..15D,1994ApJ...423..190P}. The
channel maps also clearly show a dichotomy between the NE and SW part
of the galaxy with regard to the overall intensity level of the
H$\alpha$ emission.This was already noted by
\cite{1990ApJ...352L...1R} and can be seen most clearly in the
spectacular color image of NGC 891 obtained by
\cite{1997AJ....114.2463H} (their Fig. 1) which shows a line of blue
knots all along the north side at $\vert {\rm z} \vert =0$, and no
such features on the south side. This dichotomy is also seen in the
distribution of the non-thermal radio continuum
emission. \citep{1991A&A...246...10H}

\subsection{H$\alpha$ distribution}\label{distr}
\begin{figure*}[tbp]
\centering \includegraphics[width=12cm]{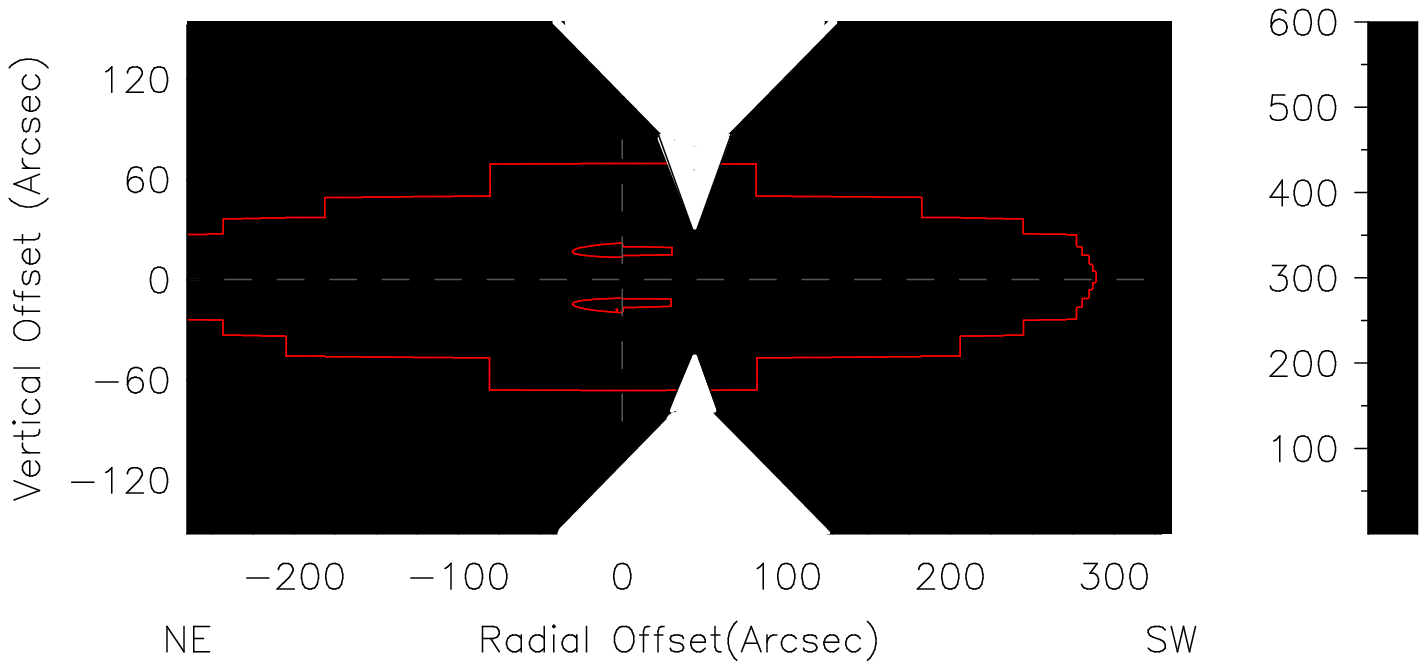}
\caption{H$\alpha$ distribution from NGC 891 as obtained from
integrating the binned channel maps. The black contours are at 1.0 and
10.0 $\times 10^{-17}$ erg s$^{-1}$ cm$^{-2}$ arcsec$^{-2}$. Red
contours are the best fit model (see $\S$ \ref{image}) The horizontal
and vertical dashed lines indicate the major and minor axis
respectively.}\label{fig4} \centering
\includegraphics[width=12cm]{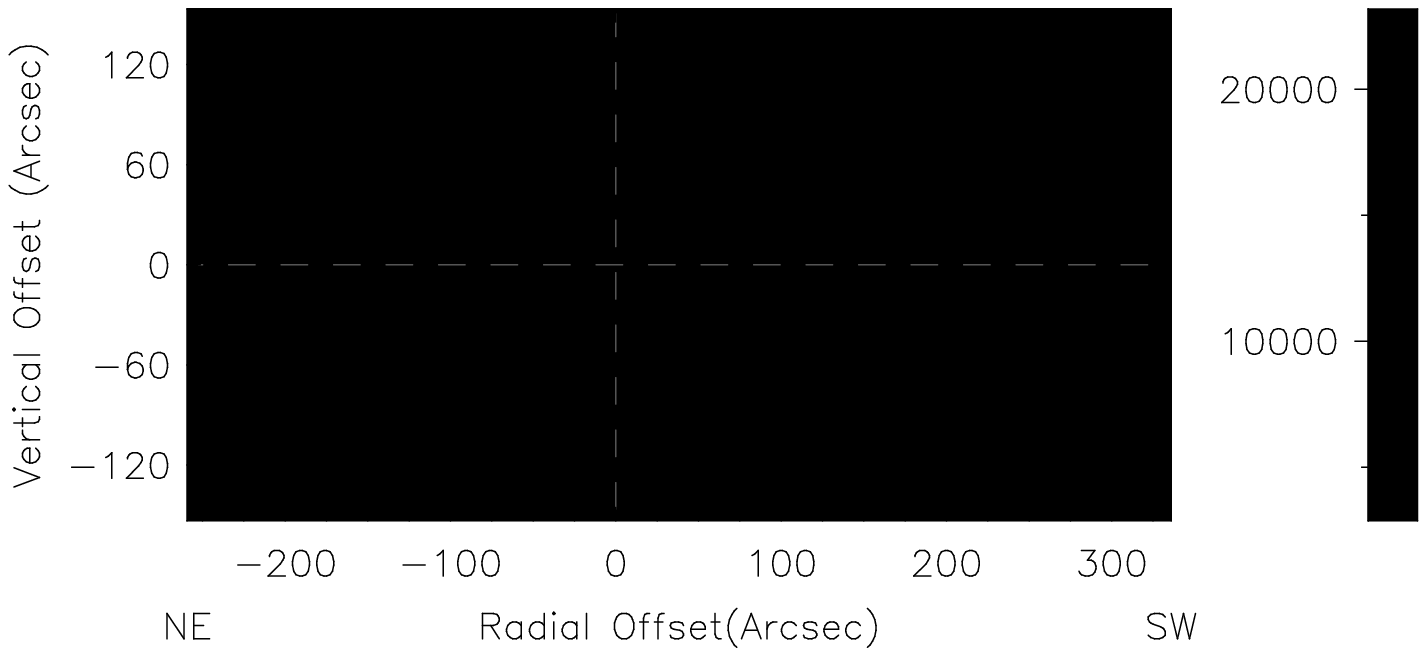}
\caption{$R$-band image taken from the Palomar Sky Survey for
comparison with the H$\alpha$ distribution.The horizontal and vertical
dashed lines indicate the major and minor axis
respectively.}\label{Rband}
\end{figure*}

The dichotomy discussed in $\S$ \ref{channel} is seen even better in
the total H$\alpha$ distribution, which is shown in Figure
\ref{fig4}. This image was obtained by integrating all channel maps
along the velocity axis of the cube. It clearly shows that the diffuse
ionized gas (DIG) extends  beyond our field of view in several places.
On the NE side of the galaxy our whole image is filled with low-level
emission; on the SW side, however, we are not able to distinguish more
than the major axis of the galaxy. This suggests that the difference
in intensity is a physical effect and not a line-of-sight effect (see
further discussion $\S$ \ref{diskz}). For comparison we added a
$F$-band image from the POSS II, below the H$\alpha$ image.
\begin{figure*}[tbp]
\centering \includegraphics[width=12cm]{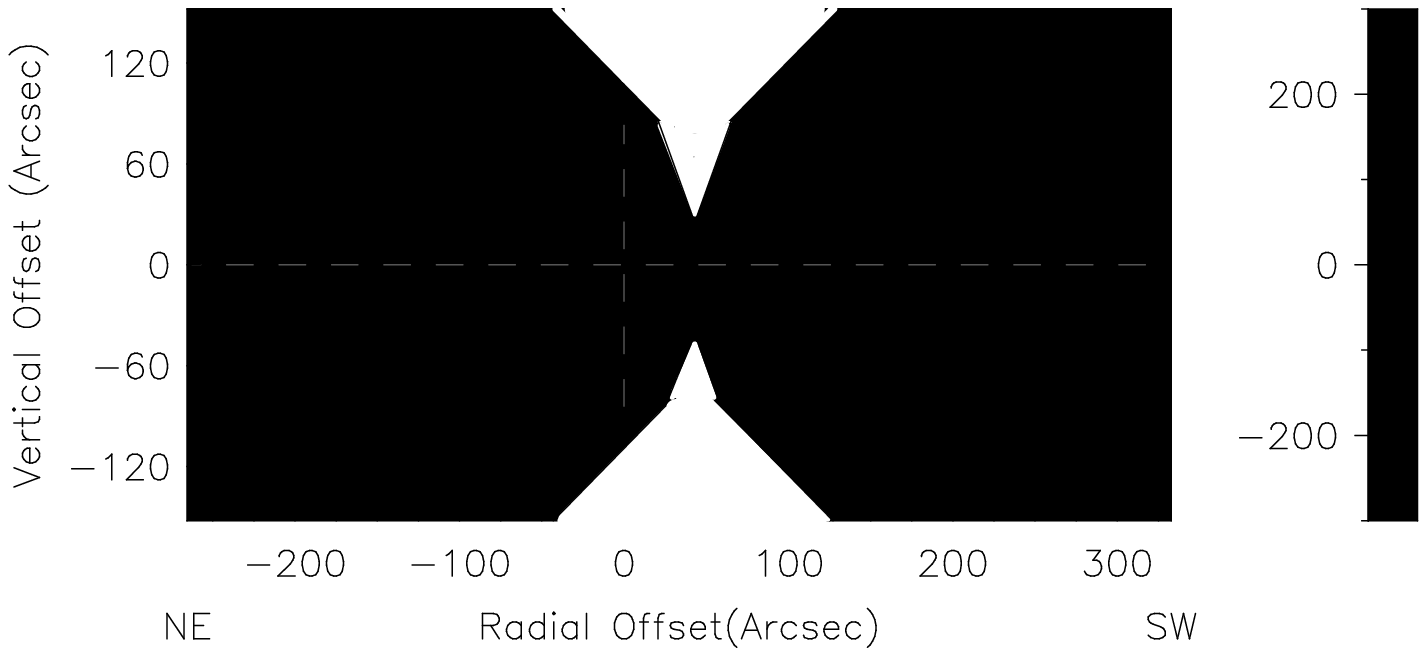}
\caption{The velocity field of the observed H$\alpha$ determined by a
Gaussian fitting to the line profiles. The contours are at -200, -100,
0, 100, 200 km s$^{-1}$ and are with respect to the systemic velocity
of 528 km s$^{-1}$. The horizontal and vertical dashed lines indicate
the major and minor axis respectively. }\label{fig5}
\end{figure*}
\subsection{Velocity field}\label{velfield}
 Fig. \ref{fig5} shows the velocity field of our H$\alpha$ cube. This
velocity field is determined by fitting a Gaussian profile to the line
profile in each bin, the peak of this Gaussian is considered to be the
velocity in this bin. This way we do not measure the real rotational
velocity but an apparent mean velocity which is determined by a
combination of the rotational velocity, the density distribution of
the gas, and the opacity of the dust. This velocity will be referred
to as the mean velocity. We chose the Gaussian fit because in the
places where the underestimation of the rotational velocity is most
significant (major axis, center of the galaxy) the H$\alpha$ is
optically thick (see discussion).\\
\hspace*{0.5 cm}On the NE side we see a regular velocity field which
resembles solid body rotation with some lower velocities in the bins
at $\vert {\rm z}\vert =60$ arcsec. This would indicate that  the
H$\alpha$ lagging does not begin below 60 arcsec (2.8 kpc). However,
as the optical depth declines we expect to look deeper into the
galaxy. This would mean that we are receiving more emission from the
line of nodes the further we are from the plane of the galaxy. Since
the real rotational velocity should be determined at the line of nodes
our underestimate of the velocity would be less the further we look
into the galaxy. So for a cylinder with a declining optical depth in
the $\vert{\rm z} \vert$-direction and solid body rotation we would
expect the mean velocities to rise as the distance from the major axis
increases. This is not the case for NGC 891 as we can see in
Fig. \ref{fig5}.\\
\hspace*{0.5 cm}If we look at Fig. \ref{fig5} and follow the -200 km
s$^{-1}$ contour we see that, starting from major axis, the  mean
velocity first rises until $\sim 20$ arcsec above the plane. Then the
mean velocity starts to drop up to $\sim 60$ arcsec. Above this the
mean velocity starts to rise again but it is unclear whether this is
real or a combined effect of the binning and Gaussian fit.\\
\hspace*{0.5 cm} Looking at the South-West side of the velocity field
we see that this side is much more irregular in velocity than the
North-East side. Following the 100 km s$^{-1}$ contour we see a
behavior similar to that of the -200 km s$^{-1}$  contour only much
more extreme. Since on the SW side above $\sim 60$ arcsec there is no
emission it is unclear if the mean velocity would start to rise again
above this height.\\
\section{Models}\label{models}
\subsection{Position - Velocity diagrams}\label{pv}
\hspace*{0.5 cm}An examination of `Position - Velocity diagrams' (PV
diagrams) provides the basis for our discussion of the
H$\alpha$. These diagrams are another representation of the channel
map data from Figs. \ref{fig3a} and \ref{fig3b}, where now the
profiles are extracted at each point along a locus of positions in the
image of the galaxy and plotted as contours in the PV plane. Figure
\ref{figXimodel} is one example of this representation; here the
position (x-axis) is measured along the major axis of the galaxy
through the nominal center at $\vert {\rm z} \vert=0$, and on the
y-axis radial velocity is given. The color scale represents the
H$\alpha$ surface brightness observed at each position; for instance,
the H$\alpha$ line profile at the position located 1 arcmin to the
North of the galaxy center would be a line parallel to the velocity
axis at a radial offset of -1 arcmin.  In the following discussion we
will concentrate on the NE side of the galaxy and refer to the
absolute velocities.
\subsection{PV-model} \label{modPV}
\begin{table}[tbp]

\begin{tabular}{llll}

\hline

Parameters & Upper limit & Lower limit & Best fit model \\     
Name & Model 1.1/M 1.2 & Model 2.1/M 2.2 & Model 3/M G3\\  
\hline

$h_{\rm g}$ (kpc)& $6.5/5.5$&3.0/2.5&5.0 \\

$z_{\rm g}$ (kpc)& 0.8/0.8&0.8/0.8& 0.8 \\

$h_{\rm d}$ (kpc)&8.1/8.1&8.1/8.1&8.1\\

$z_{\rm d}$ (kpc)&0.26/0.26&0.26/0.26&0.26\\

$\tau_{{\rm H \alpha}}$ &4-6/6 &12-14/13-14& 6\\

$\sigma_{\rm v}$ (km s$^{-1}$)&40/40&40/40&40\\ $R_{\rm
max}$&21/14&21/14&14\\

\hline

\end{tabular}

\caption{Model parameters. The addition G to a model indicates that
the rotation curve has a gradient of -18.8 km s$^{-1}$ kpc$^{-1}$ in
the vertical direction. For the upper and lower limit there are two
models which differ in truncation radius (See $\S$ \ref{modPV} ).
$h_{\rm g}$ and $z_{\rm g}$ are the scale length and height of the gas
respectively. $R_{\rm max}$ is the truncation radius of the models.}
\label{tab2}
\end{table}

\begin{figure}[tbp]
\centering \includegraphics[angle=0 ,width=8cm]{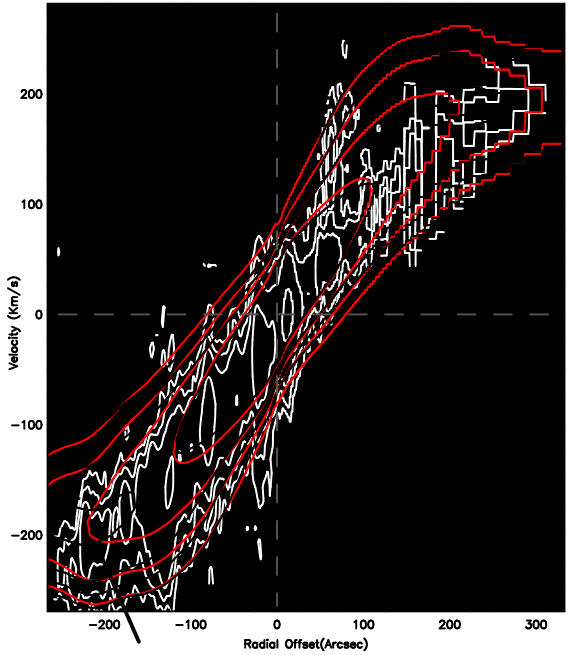}
\caption{Color plot of the H$\alpha$ PV-diagram along the major axis
overlaid with contours of Model 1 and 2 (see Table \ref{tab2})
corresponding to the upper and lower limit which can be fitted. White
contours are the data at 3$\sigma$, 6$\sigma$, 12$\sigma$ and
48$\sigma$. Black contours are Model 1.1 with $\tau_{\rm H \alpha}=5$,
red contours are Model 2.1 with $\tau_{\rm H \alpha}=13$. The black
arrow indicates the place where the fits deviate from the data (see
text).}\label{figXimodel}
\end{figure}

\begin{figure}[tbp]
\centering \includegraphics[angle=0 ,width=8cm]{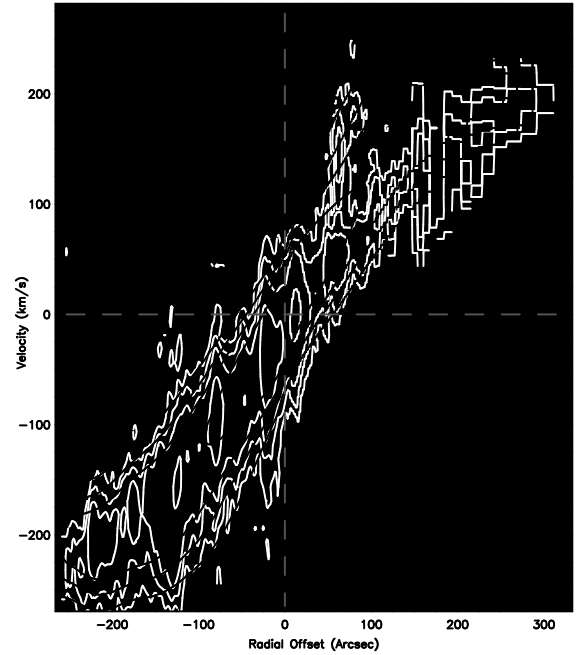}
\caption{Color plot of the H$\alpha$ PV-diagram along the major axis
overlaid with the contours of the best fit model.White contours are
the data at 3$\sigma$ 6$\sigma$ 12$\sigma$ 48$\sigma$.  Black solid
contours are Model 3 (Table \ref{tab2}). }\label{fig9}
\end{figure}

\hspace*{0.5 cm} The position velocity model is a FORTRAN code which
calculates emission at every position of an exponential disk taking
line of sight velocities into account. For every position the light is
extincted as expected from a dust disk with a given optical depth and
an exponential distribution with variable scale length and height. The
structural parameters are defined in the same way as in the models
used by \cite{1998A&A...331..894X}. If the radius of the disk exceeds
a certain cut off ($R_{\rm max}$) all emission and absorption is set
to 0. This is done to simulate a truncation radius. The code then
integrates these values along the line of sight at every position and
determines an observed velocity distribution and a
intensity. Scattering is ignored in the calculations.\\
\hspace*{0.5cm}The code allows the disk to be inclined and for NGC 891
we chose an inclination of 89$^{\circ}$.  To make a fit we assumed the
HI rotation curve \citep{2005ASPC..331..239F} and  a truncation radius
$R_{\rm max}=21$ kpc \citep{1981A&A....95..116V}. We fit the NE (left)
side of the PV-diagram by eye. The SW (right) side is not taken into
account because the signal is very irregular on this side (see $\S$
\ref{velfield}). We started with fitting some simple models where the
scale length of the dust equals the scale length of the gas ($h_{\rm
g}=h_{\rm d}=5$ kpc).\\
\hspace*{0.5 cm}These simple models all show a major discrepancy with
the data at high velocities and large radii, where the intensities in
the models start to rise again while in the data no such rise is seen.
To solve this problem we needed the scale length of the dust to be
longer than the scale length of the gas. Therefore we assumed that the
dust disk has a scale length of 8.1 kpc \citep{1998A&A...331..894X}.\\
\hspace*{0.5 cm}   With this longer scale length for the dust the
intensity peak at large radii and high velocities  has disappeared and
the general shape of the PV-diagram is now comparable to the data.
Also it provides us with an upper limit for the scale length of the
gas. At a gas scale length of 6.5 kpc (Model 1.1, see Table
\ref{tab2}, Fig. \ref{figXimodel} black contours) the problem of the
simple model arises again. This is seen in Fig.  \ref{figXimodel}
around -200 arcsec and -200 km s$^{-1}$ offset (pointed out by the
black arrow) where the highest black contour of $48 \sigma$ reappears
while in the data no such thing is seen. Therefore we consider Model
1.1 as a upper limit for the gas scalelength. A lower limit for the
scale length is found at 3.0 kpc (Model 2.1, see Table \ref{tab2},
Fig. \ref{figXimodel} red contours). At this scale length we clearly
see the second highest contour bending up around an offset of -200
arcsec and -200 km s$^{-1}$ (pointed out by the black arrow) while the
same contour for the data continues almost up to the edge of the image
at an offset of -230 arcsec and -225 km s$^{-1}$. Figure
\ref{figXimodel} also shows that we are overestimating the intensities
at low velocities.  To fit the low velocities at large radii a shorter
truncation radius is needed in the models. Unfortunately this
truncation radius is clearly outside our field of view. Thus we can
only find the right truncation by fitting the PV-diagram. We find that
a truncation at 14 kpc (${\rm R_{max}}$) fits the data the best. This
differs from the radius of the optical truncation (${\rm R_{max}}$)
determined by \cite{1981A&A....95..116V} who obtain ${\rm R_{max}}$=
450 arcsec (21 kpc) but is in agreement with
\cite{1990ApJ...352L...1R} who find diffuse emission out to 15 kpc.
When we determine the upper and lower limit on the scale length for
models with a the new truncation radius for the ionized gas (14 kpc)
(Model 1.2 and Model 2.2, see Table \ref{tab2})we find that these
models show the same behavior as Model 1.1 and Model 2.2  but at
shorter scale lengths. This is caused by the fact that in the models
not only the gas disk is now truncated at 14 kpc but the dust disk as
well. It remains unknown which truncation is more suitable for the
dust disk.\\
\hspace*{0.5 cm} \cite{2006AJ....131..716K} recently found that on
average the H$\alpha$ scale length for a field galaxy is on average 14
\% longer than the stellar scale length.  Based upon the value found
by \cite{1998A&A...331..894X} in the $V$-band this would mean that the
deprojected H$\alpha$ scale length for NGC 891 should be $\sim$ 6.5
kpc which is in agreement with our limits.\\
\hspace*{0.5cm}Although the central attenuation for a given scale
length is quite well constrained, the differences in dust attenuation
can be quite large between the different scale lengths. This gives us
another handle on which scale length is
correct. \cite{1998A&A...331..894X} found a central optical depth of
$\tau_{\rm face-on}=0.7 \pm 0.01$ in $V$-band, for the galaxy seen
face on. For our models for this edge-on galaxy this would translate
to a central attenuation of $\tau_{\rm H \alpha}=10.9$. $\tau_{\rm H
\alpha}$ in our models is the optical depth at a radial and vertical
offset of 0 along the line of sight to the center of the disk.\\ We
consider the model with $h_{\rm d}=5.0$ kpc, $\tau_{\rm H \alpha}=6$
and $R_{\rm max}$=14 kpc (Model 3) the best fit. Fig. \ref{fig9} is an
example of the major axis PV diagram of the data overlaid with
contours of Model 3. Given the dependence of the central optical depth
on scale length our results are not in disagreement with
\cite{1998A&A...331..894X}.\\
\subsection{Image-model} \label{image}
\begin{figure}[tbp]
\centering \includegraphics[angle=90,width=8cm]{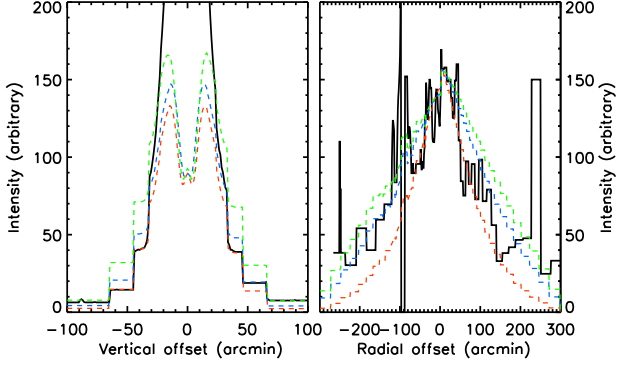}
\caption{Normalized intensity cuts of integrated maps. Left: along the
minor axis. H$\alpha$ (solid line)), $z_{\rm g}$=0.7 kpc (dashed red
line), $z_{\rm g}$=0.8 kpc (dashed blue line), $z_{\rm g}$=0.9 kpc
(dashed green line). Right: Parallel to the major axis at a vertical
offset of 30 arcsec. H$\alpha$ (solid line)), $h_{\rm g}$=3.0 kpc
(dashed red line), $h_{\rm g}$=5.0 kpc (dashed blue line), $h_{\rm
g}$=6.5 kpc (dashed green line) }\label{intcut}
\end{figure}

After we fitted the PV-diagram on the major axis we put the same
values into a FORTRAN code which calculates an intensity along the
line of sight (see $\S$ \ref{pv}). This code produces a model image
which we can compare with the observed images of NGC 891. To determine
the correct scale height we compare an intensity cut parallel to the
minor axis averaged between -100 to -50 arcsec of the model images to
the observed H$\alpha$ distribution averaged over the same region
(Fig. \ref{fig4}). Since at this point we are interested only in the
vertical shape above the dust, the maps are first normalized to their
emission 30 arcsec above the plane. To determine the best fit we
concentrate on the emission at a positive offset of the plane since
this side is brightest. In our fit we only consider the emission at
offsets larger than 30 arcsec. From this comparison we find that a
scale height of 0.8 kpc best fits the data. We then determine from
this comparison a scaling for the model so that it represents the
unnormalized data. Figure \ref{intcut} (left) shows this averaged  cut
along the minor axis. This figure shows the data (solid line) and the
scaled model for $z_{\rm g}$=0.7 (dashed red line), 0.8 (dashed blue
line), 0.9 (dashed green line) kpc. We see that $z_{\rm g}$=0.8 kpc is
the best fit to the data.\\
\hspace*{0.5cm} As a check on our scaling factor and our scale lengths
Fig. \ref{intcut} shows on the right a cut parallel to the major axis
at a vertical offset of 30 arcsec. The solid black line is the data
and the colored lines are the scaled models with a changing scale
length with $h_{\rm g}$=3.0 kpc (dashed red line), $h_{\rm g}$=5.0 kpc
(dashed blue line) and $h_{\rm g}$=6.5 kpc (dashed green line).  This
figure shows clearly that a scale length of 5 kpc is the best fit to
the data.
\subsection{Cube model} \label{modcube}
Having obtained the best fits for the images and the major axis
PV-diagram we model a full data cube so we can obtain PV-diagrams at
any height in the disk. We constructed two of these cubes based on the
the best fit of the major axis PV-diagram. These cubes are then binned
in the same way as the data  and scaled with the previously derived
scaling factor. In one of these cubes the rotation curve is kept
constant throughout the vertical distribution of the cube (Model 3,
see Table \ref{tab2}). The other cube model contains a vertical
gradient for the rotation curve of -18.35 km s$^{-1}$ kpc$^{-1}$
(Model G3, see Table \ref{tab2}). In this model the radial shape of
the rotation curve is not changed. These cubes and their comparison to
the data will be presented below.
\section{Discussion}\label{disk}

\subsection{Kinematics in the plane}\label{diskplane}
\begin{figure}[tbp]
\centering \includegraphics[width=8cm]{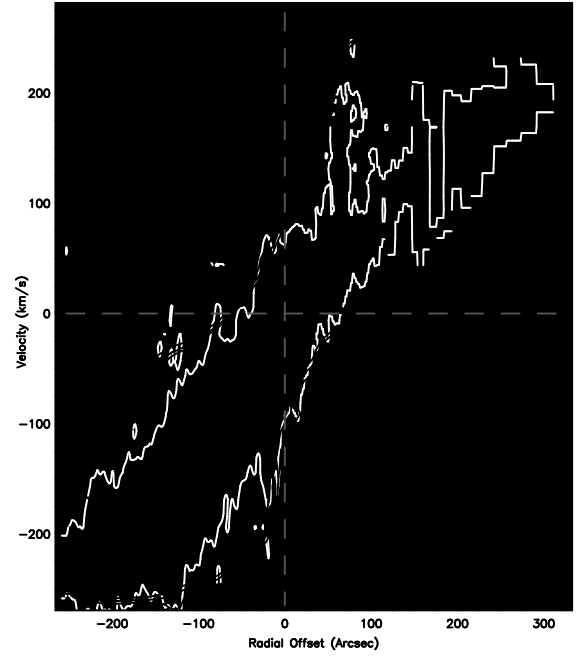}
\caption{Color plot of the H$\alpha$ PV diagram along the major
axis. For comparison the HI PV diagram on the major axis is over
plotted at contour levels 3$\sigma$, 6$\sigma$ and 9$\sigma$.
}\label{fig12}
\end{figure}
Figure \ref{fig12} shows a PV diagram of the  H$\alpha$ emission along
the major axis of the galaxy ($\vert z \vert=0$). This diagram bears
the signature of solid body rotation instead of showing the strong
differential rotation of the HI. The simplest interpretation of this
is that the disk of the galaxy is optically thick at $\vert z
\vert=0$, so that the H$\alpha$ emission we see is mostly coming from
the front edge of the disk. This is consistent with $\tau_{\rm H
\alpha}$=6.\\
\begin{figure}[tbp]
\centering \includegraphics[angle=90,width=8cm]{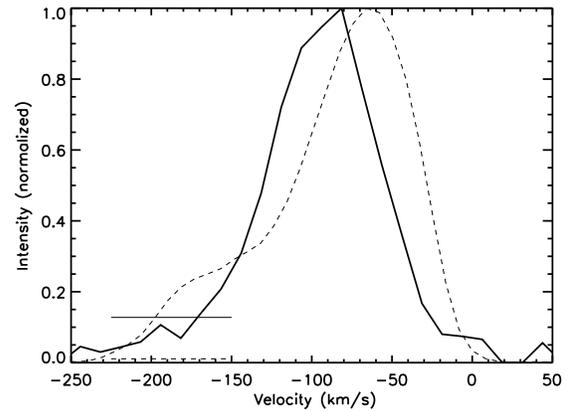}
\caption{Normalized velocity line profiles of the H$\alpha$ (Solid
line) and HI(dashed line)on the major axis  at a radial offset of -1
arcmin. The straight lines indicate the 3 $\sigma$ values for both
observations}\label{Specplot}
\end{figure}
\hspace*{0.5 cm}Let us consider one `cut' through this diagram
parallel to the velocity axis, at a radial offset of -1 arcmin
(Fig. \ref{Specplot}). Presuming that the H$\alpha$ emission emanates
from gas which is in circular rotation, the H$\alpha$ emission at
$\vert 25 \vert$ km s$^{-1}$ is at the very front of the disk.  There
is an absence of emission at lower velocities because the H$\alpha$
disappears as we get to the front edge of the disk of the galaxy. As
we descend in this diagram towards $\vert 175 \vert$ km s$^{-1}$, at
the same radial offset, the emission fades out. We interpret this as a
result of increasing extinction due to dust in the plane. From our
best fit model, approximately 6.5 magnitudes of extinction, along the
line of sight to the center of the galaxy, are implied by this
interpretation of the data; assuming that there is no extinction in
the HI.\\
\hspace*{0.5 cm} As we sample H$\alpha$ emission at larger radial
offset we look closer to the line of nodes, and the  velocities
increase until we actually look at the line of nodes and the
velocities do not rise anymore. Note though that due to the clumpiness
of the emission sources the velocities can still decrease after this
point.  \\
\hspace*{0.5 cm}Alternatively, the H$\alpha$ emission may be confined
to a  thin annulus in the galaxy. This annulus would have to be in the
outer  parts of the galaxy, with little or no emission inside it.  We
consider such a distribution  of the H$\alpha$ to be unlikely,
especially in the view of the H$\alpha$ at higher $\vert {\rm
z}\vert$, as we shall discuss in section \ref{diskz}.\\
\hspace*{0.5 cm} Figure \ref{fig12} clearly shows the dichotomy
between the NE and SW discussed earlier (sect. \ref{result}). If NGC
891 has spiral arms, the asymmetry suggests that the H$\alpha$
emission on the north side is emanating from H{\small II} regions
located on the outside of the spiral arm, while to the south we are
viewing the opposite arm from the inside. This suggested morphology is
also consistent with the fact that the North-East side of the galaxy
is approaching us, while the South-West side is receding, since then
the spiral arms are trailing. From the ratio of emission between the
NE and the SW side along the along the major axis this morphology
implies an extra 1.1 magnitudes of extinction on the SW side due to
the spiral arm. \\
\subsection{Kinematics at high z}\label{diskz} 
\begin{figure}[tbp]
\centering \includegraphics[angle=0,width=8cm]{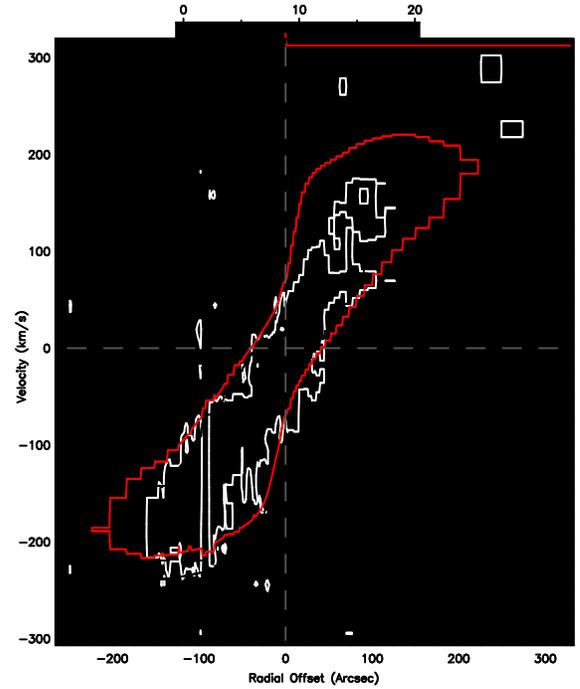}
\caption{Color plot of the H$\alpha$ PV-diagram at 26-34 arcsec
  (1.2-1.6 kpc) offset from the major axis. Contours are at 3
  $\sigma$. The white solid contour is the data, black contour is the
  best fit model (Model 3, Table\ref{tab2}), red contour is the best
  fit model with an assumed vertical gradient of 16.5 km s$^{-1}$
  kpc$^{-1}$ in the rotation curve (Model G3), Table
  \ref{tab2}.}\label{fig10}
\end{figure}

\begin{figure}[tbp]
\centering \includegraphics[angle=0,width=8cm]{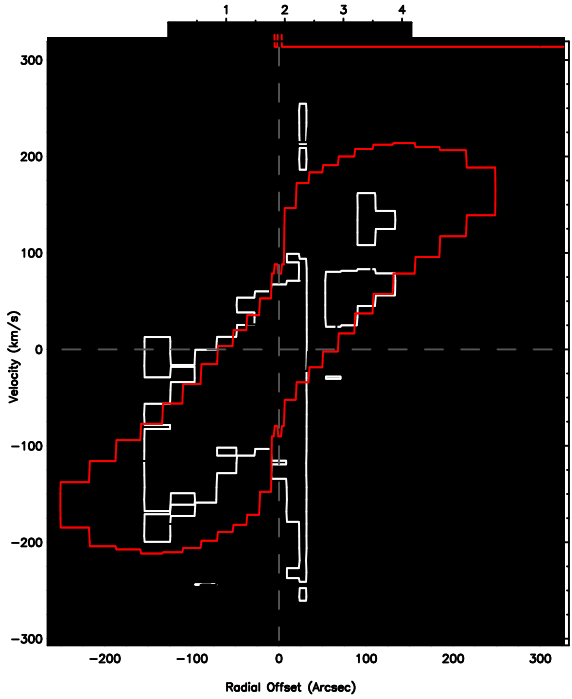}
\caption{Same as Fig. \ref{fig10} except now for the bin 48-67 arcsec
  (2.2-3.1 kpc) offset from the major axis.}\label{fig11}
\end{figure}

\begin{figure}[tbp]
\centering \includegraphics[angle=0,width=8cm]{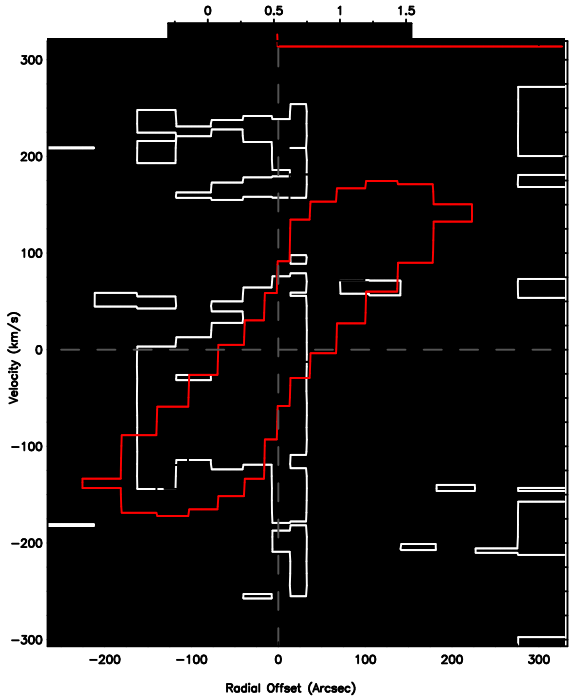}
\caption{Same as Fig. \ref{fig10} except now for the bin 68-107 arcsec
  (3.2-4.9 kpc) offset from the major axis.}\label{arc90}
\end{figure}

\begin{figure}[tbp] 
\centering  \includegraphics[angle=90,width=8cm]{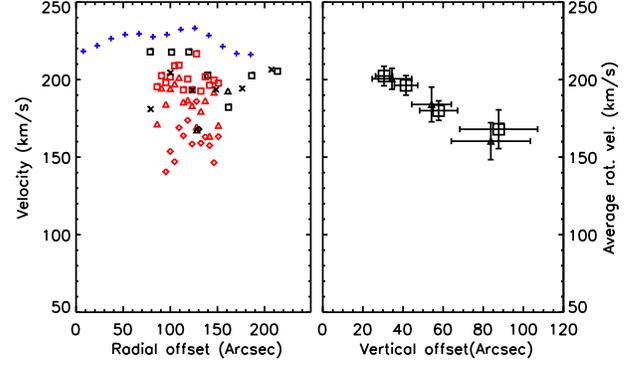}
\caption{(Left) Results of employing envelope tracing to the
data. Blue (pluses): HI on the major axis \citep{2005ASPC..331..239F},
red: points found by \cite{2006ApJ...647.1018H},  at vertical offsets
$\sim$ 30 arcsec (squares), $\sim$ 50 arcsec (triangles) and $\sim$ 80
arcsec (diamonds). Black: our data at vertical offsets of 30 arcsec
(squares), 40 arcsec (crosses), 58 arcsec (triangles) and 88 arcsec
(diamonds).  (Right) The average rotational velocity obtained with
envelope tracing at a vertical offset of 30, 40, 58 and 88 arcsec
(1.4, 1.9, 2.7 and 4.1 kpc) for our data (squares) and average points
as obtained by \cite{2006ApJ...647.1018H} at vertical offsets $\sim$
30, 50, 80 arcsec (1.4, 2.4, 3.7 kpc)(triangles).}\label{gradient}
\end{figure}

Figures \ref{fig10}, \ref{fig11} and \ref{arc90} show velocity cuts
parallel to the major axis at an offset of 24-33, 46-65 and 66-104
arcsec respectively.\\
\hspace*{0.5cm}The first thing that we notice from these figures is
that the dichotomy in intensity is also clearly visible above the
plane. In fact, as we can see from Figs. \ref{fig11} and \ref{arc90},
above $\vert {\rm z} \vert \sim 30$ there is not enough emission on
the SW side to say anything sensible about the rotation of the gas.\\
\hspace*{0.5 cm} Since dust absorption above the plane is likely to be
negligible this fact suggests that the dichotomy is a real physical
effect and that star formation in the SW is less intense, assuming the
extra-planar gas is indeed brought up from the disk by a mechanism
related to star formation.\\
\hspace*{0.5 cm} As an initial guess of the gradient, and to compare
with the observations of \cite{2006ApJ...647.1018H}, we performed
envelope tracing on Fig. \ref{fig10}, Fig. \ref{fig11} and
Fig. \ref{arc90}. Envelope tracing basically fits Gaussian profiles,
with a dispersion equal to the intrinsic dispersion of the gas
convolved with the instrumental dispersion, to the three points with
the highest rotational velocity above 3$\sigma$. The peak position of
the fitted Gauss is then considered the rotational velocity.  This
method is not very trust worthy above the plane of the galaxy where
the S/N can become low (see \cite{2005ASPC..331..239F})\\
\hspace*{0.5 cm} The points obtained with this method are shown in
Fig. \ref{gradient} (left). For comparison, Figure \ref{gradient}
(left) also shows the HI rotation curve on the major axis and the
results of \cite{2006ApJ...647.1018H}. We see that in general our data
is in agreement with their SPARSE-PAK observations. Since we have a
full cube  we can study the slope of the rotation curve in the inner
parts. We find that above the plane the rotation curve rises less
steeply with radius the further we get from the plane. The HI
observations already hinted at this but due to the resolution this
result could not be confirmed. At every height we average the points
obtained at radii larger than 80 arcsec. These points are shown in
Fig. \ref{gradient} (right). With these three points we find from
envelope tracing a gradient of 15 $\pm$ 6.3 km s$^{-1}$ kpc$^{-1}$ .\\
\hspace*{0.5 cm} Figure \ref{fig10} shows that the general slope of
the diagram steepens compared to PV-diagram at the major axis. This is
as we would expect since the gas is less obscured by the dust above
the plane. Therefore, we can look farther into the galaxy and look at
gas closer to the line of nodes. This steepening is also the reason
why a thin annulus in the outer parts of the galaxy (See $\S$
\ref{diskplane}) is very unlikely. In such a distribution this
steepening would not be possible unless the gas of the annulus would
move inward as it rises above the plane. Such an effect seems highly
unlikely.\\
\hspace*{0.5 cm}If we compare the data to  Model 3 we see that the
steepening is not enough. Our model has much more gas at high
velocities near the center of the galaxy. This lack of gas at the high
velocities in the center might still be an effect of the dust but
could also indicate that the rotational velocities of the gas above
the plane not only lag compared to the disk but that the  rotation
curve rises less steep radially the higher we look above the plane.\\
\hspace*{0.5 cm}A close inspection of Figure \ref{fig10} shows us that
there are two more places where the data deviate from the model.  The
model underestimates the intensities at low velocities and
overestimates them at high  velocities.  The lack of gas at high
velocities at all radii confirms the lagging rotation curve found by
\cite{2005ASPC..331..239F} and \cite{2006ApJ...647.1018H}. If we draw
a straight line through the lower part of the 3$\sigma$ contour of the
data and and then draw a straight line through the same contour of
Model 3 we can measure the lagging of the halo. In this way we find a
difference between Model 3 and the data $\sim 18.75 \pm 6.3$ km
s$^{-1}$ at a vertical offset of 30 arcsec (1.4 kpc).\\
\hspace*{0.5 cm}In the diagram that shows the gas at an offset of 60
arcsec (Fig. \ref{fig11}) we see that  the slope of the emission
becomes less steep compared to the slope at 30 arcsec. This is the
continued effect of the rotation curve rising less steeply with radius
the further we get from the plane. For this effect to be caused by
dust the dust extinction would have to increase again which seems
highly unlikely.\\
\hspace*{0.5cm} From Figure \ref{fig11} we find a difference between
Model 3 and the data, by comparing the 3$\sigma$ contours, of $62.5
\pm 6.3$ km s$^{-1}$. At this height we cannot be completely certain
we are looking at the flat part of the rotation curve. Therefore,
these effects could also be caused by radial redistribution of the
gas. We consider it unlikely that such a redistribution completely
causes the changes of the observed PV-diagram because intensity cuts
parallel to the major axis only show a hint of such an effect and only
at the East side of the galaxy, as shown by \cite{2006ApJ...647.1018H}
(their Fig. 7), while the West side is the brighter side of the
halo. \\
\hspace*{0.5 cm} Figure \ref{arc90} shows the gas at 90 arcsec offset
from the major axis. The emission of the diffuse gas at this height is
very low and we had too compare the 1$\sigma$ contours of the model
and the data. Therefore conclusions drawn from this plot are
considered to be no more than indicative. At this vertical offset the
effects observed at a 60 arcsec offset continue. Comparing the highest
velocity of the 1$\sigma$ contour at this height with Model 3 we
observe a difference of $81.25 \pm 12.5$ km s$^{-1}$.\\
\hspace*{0.5cm}When we assume that the gradient starts on the  major
axis we find the slope of the gradient to be $\sim 18.8 \pm 6.3$ km
s$^{-1}$ when we fit the points at 30, 60 and 90 arcsec  (1.4, 2.7 and
4.1 kpc).\\
\hspace*{0.5 cm} After determining the gradient of the lag we
constructed a model (Model G3, see Table \ref{tab2}) in which the
rotation curve is scaled down at higher $\vert z \vert$ by subtracting
at every vertical step in the model $\vert z \vert \times 18.8$ km
s$^{-1}$, with $\vert z \vert$ in kpc,  from the rotation curve as
obtained from the HI. The vertical step size in the model was 49 pc
(1.05 arcsec). Model G3 is plotted in Figs. \ref{fig10}, \ref{fig11}
and \ref{arc90} as the red contours. We see that gas is still missing
at various places in the diagram but that the maximum  and minimum
velocities are  approximately the same for the data as this model at
the 3$\sigma$ contour. Thus confirming that there is a gradient of
-18.8$\pm$6.3 km s$^{-1}$ kpc$^{-1}$ in the observations. The
explanation for the missing gas remains the same as before since we
did not change the shape of the rotation curve.\\
\begin{figure}[tbp] 
\centering \includegraphics[angle=0,width=8cm]{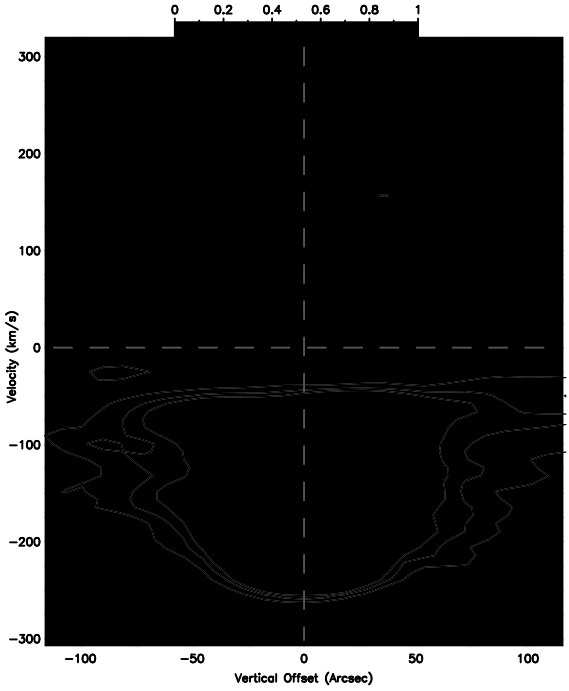}
\caption{Normalized color plot of the H$\alpha$ PV-diagram at a radial
offset of 150 arcsec parallel to the minor axis and overlaid with
contours of HI at 3$\sigma$, 6$\sigma$ and 9$\sigma$.}\label{figlog150}
\end{figure} 
\begin{figure}[tbp]
\centering  \includegraphics[angle=0 ,width=8cm]{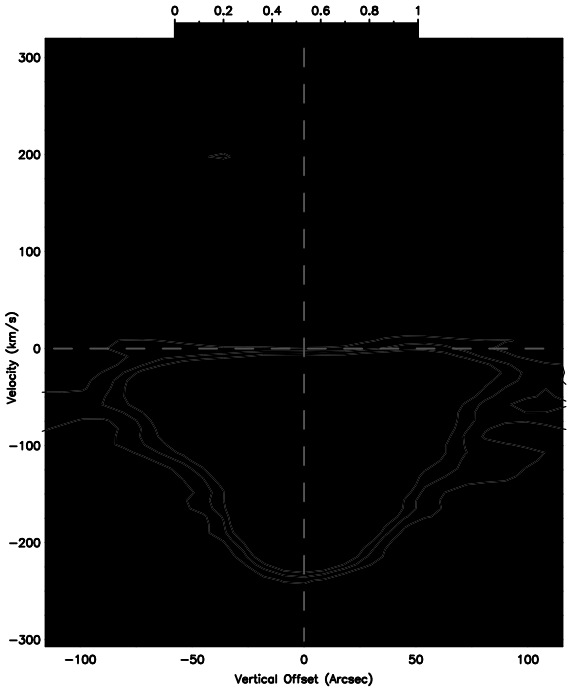}
\caption{Same as \ref{figlog150} but now at a radial offset of 75
arcsec.}\label{figlog75}
\end{figure} 
\begin{figure}[tbp] 
\centering \includegraphics[angle=0 ,width=8cm]{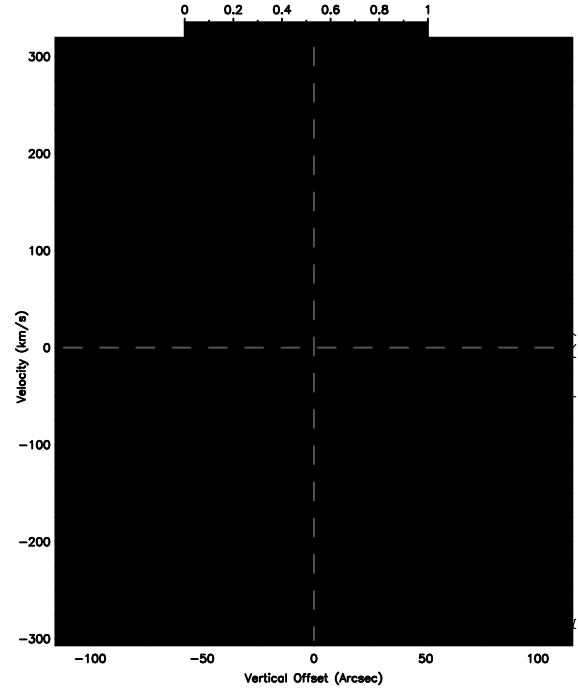}
\caption{Same as \ref{figlog150} but now at the minor axis
position.}\label{figlogmin}
\end{figure}
\hspace*{0.5cm} Another way to look at the kinematics at higher $
\vert z \vert$ is by constructing PV diagrams along the minor axis and
parallel to the minor axis at some radial offset. To optimize the
information in the diagrams we normalized them by dividing every line
profile by it's maximum. Figure \ref{figlog150} is an example of such
a PV-diagram. This PV-diagram is constructed by looking through the
cube at a radial offset of 150 arcsec and is a cut parallel to the
minor axis. Overlaid on the color scale are the 3, 6, 9 $\sigma$
contours of the HI. Looking at this plot the first thing we see is
that the HI is much more extended vertically than the H$\alpha$. This
is partly due to beam smearing but not completely. If we look at the
H$\alpha$ at low mean velocity ($\sim  \vert 140 \vert$ km s$^{-1}$)
we see that in the plane of the galaxy (e.g. 0 offset) the maximum of
the emission lies at this  low mean velocity. Moving away from the
plane the maximum of the emission first rises to  higher mean
velocities and then drops again. The initial rise is caused by
diminishing  dust attenuation. As we move further from the plane the
maximum of the emission drops to lower mean velocities again. This
drop is caused by the lower rotational velocities at higher $\vert z
\vert$.  The H$\alpha$ is much less extended than the HI, in velocity
as well as vertical size ($z_{\rm EM}$=0.5 kpc
\citep{1990A&A...232L..15D}, the ionized gas scale height is twice
this, $z_{\rm HI}$=2.3 kpc (T.Oosterloo, priv. communication)).
Considering the sensitivity of both observations it could well be that
our H$\alpha$ observations are just not sensitive enough to observe
all the ionized gas in the galaxy. We also plot PV diagrams parallel
to the minor axis at a offset of 75 arcsec and on the minor axis
itself, Figures \ref{figlog75} and \ref{figlogmin} respectively.  In
the figure at 75 arcsec offset from the minor axis we see the same
behavior as at 150 arcsec offset, only here the rise and drop in  mean
velocities is much more extreme. We also see that in this diagram the
mean velocity at the highest offset from the plane drops back towards
systemic. This difference is caused by the rotation curve which rises
less steep the further we look above the plane. Looking at the diagram
which is a cut along the minor axis of the galaxy
(Fig. \ref{figlogmin}) we see that here the maximum of the H$\alpha$
emission lies one channel below systemic velocity at almost all the
offsets from the plane. This offset is about 18 km s$^{-1}$ which is
larger than the error in the wavelength calibration (e.g., 6 km
s$^{-1}$). We are confident this offset is not an error in our
velocity scale. In principle we could check this by comparing the flat
rotation speeds on the North side to those on the South side but we
think such a check is unreliable due to the effects of dust on the
South side.\\
\hspace*{0.5 cm}The flat shape in Figure \ref{figlogmin} is as
expected; the offset from systemic is unexpected. We realize that our
central position of the galaxy is somewhat to the south from the
central position generally used in kinematical studies, but notice
that shifting the central position to the north would further remove
us from the kinematical center of the H$\alpha$. Also our kinematical
center and the center used in this paper would lie in the same
resolution element of the HI observations.  We note that at a vertical
offset of $\sim$ 60 arcsec the maximum seems to be displaced more
from the systemic velocity. This is a real effect and is not caused by
our way of binning the data. Whether this deviation is important for
understanding the general dynamics of the halo remains unclear.
\section{Summary}\label{summary}
\hspace*{0.5 cm}We present Fabry-P\'erot H$\alpha$ measurements of the
edge-on galaxy NGC 891. This is the first time kinematical data for
the H$\alpha$ are presented for the whole of NGC 891.\\
\hspace*{0.5 cm}In our observations we can clearly see H$\alpha$
emission above and below the plane of NGC 891. This vertical extent is
already visible in the separate channel maps and becomes even more
obvious in a velocity integrated map.\\
\hspace*{0.5 cm} This integrated velocity map  shows a clear contrast
between the distribution of the H$\alpha$ on the North-East and the
South-West side of the galaxy. This dichotomy is not restricted to the
plane of the galaxy but is also clearly visible above the plane.
Since dust absorption is negligible above the plane it is likely that
this dichotomy is a real physical effect. Assuming that the halo gas
is brought up from the plane by a SFR related mechanism, this implies
that the SFR on the South-West side of the galaxy is much lower than
on the North-East side of the galaxy.\\
\hspace*{0.5 cm}For the interpretation of the kinematics of the
extra-planar gas we constructed several 3-D models of an exponential
disk rotating with a rotation curve derived from the HI data
\citep{2005ASPC..331..239F}. Included in the models is a uniform dust
layer of given optical depth distributed exponentially in radius and
height and a truncation radius.\\
\hspace*{0.5 cm}We started with models that have the same scale length
 for the dust disk as the H$\alpha$ disk ($h_{\rm g}=h_{\rm d}=5$
 kpc). We find that such models generate too much intensity at large
 radii and high velocities when we compare them to the data.  To
 overcome this problem we modeled the galaxy with a dust scale length
 of 8.1 kpc, as derived by \cite{1998A&A...331..894X} from
 observations in the V-band. The longer scale length of the dust
 reduces the intensity of the gas at large radii and high velocities.
 This also provides us with a upper limit scale length of the ionized
 gas of 6.5 kpc (Model 1.1). Longer scale lengths would reintroduce
 the too high intensities found in the first models. A lower limit is
 found for a model with a scale length of 2.5 kpc (Model 2.2). Models
 with even shorter scale lengths do not produce enough intensity at
 large radii. Better constrains could be obtained if the truncation
 radius of the dust disk would be known.\\
\hspace*{0.5 cm} When we fit models in this range to the PV-diagram of
the major axis we find that the best fit is a model with a central
attenuation of $\tau_{\rm H \alpha}=6$, a cut off radius $R_{\rm
max}=14$ kpc and a scale length and height of 5.0 kpc and 0.8 kpc
respectively (Model 3).  By comparing PV-diagrams above the plane to
the models kinematical information about the galaxy is extracted from
the data. We confirm the lagging of the halo,  as found by
\cite{2005ASPC..331..239F} and \cite{2006ApJ...647.1018H}, and
determine that this lagging occurs with gradient of $\sim 18.8 \pm
6.3$ km s$^{-1}$ kpc$^{-1}$.\\
\hspace*{0.5 cm} In the PV-diagrams we also see that compared to the
models the distribution of the H$\alpha$ is displaced to larger radii
or lower rotational velocities. This effect increases as we look
higher above the plane. This means that the higher we look above the
plane, the less steep the rotation curve rises. We can confirm this by
comparing three cuts through the cube along and parallel to the minor
axis. After normalizing these PV-diagrams we can clearly see that the
H$\alpha$ at a distance of 75 arcsec from the center has a larger
gradient than the H$\alpha$ at 150 arcsec from the center. \\
\begin{acknowledgements}
We wish to thank the referee R. Rand for many useful comments,
F.Fraternali for providing the HI rotation curve, T. Oosterloo for
providing the HI data on NGC 891, G. Heald  and R. Rand for providing
their H${\alpha}$ rotation points and a calibrated H$\alpha$ image,
M. Potter for providing the DSS positions for the stars used for the
astrometry, and R. Sancisi for insightful comments and discussion on
the paper.
\end{acknowledgements}

\bibliographystyle{aa-package/bibtex/aa}   \bibliography{ref}
\end{document}